\pdfoutput=1
\documentclass[11pt]{article}

\usepackage{booktabs} % 用于三线表
\usepackage{array} % 用于自定义列对齐
\usepackage[caption=false]{subfig} 

\usepackage{amsmath}

\usepackage[review]{acl}
\usepackage{times}
\usepackage{latexsym}

\usepackage{booktabs} % 用于更美观的表格线
\usepackage{adjustbox} % 用于调整表格宽度
\usepackage{array} % 提供更多表格对齐选项
\usepackage{geometry} % 用于调整页面布局

\usepackage[T1]{fontenc}
\usepackage[utf8]{inputenc}

\usepackage{microtype}

\usepackage{inconsolata}

\usepackage{graphicx}

\title{Simulating Human Behavior with the Psychological-mechanism Agent: Integrating Feeling, Thought, and Action}

\author{Qing Dong, Pengyuan Liu, Dong Yu, Chen Kang \\
    Beijing Language and Culture University, Beijing, China}

\begin{document}
\maketitle
\begin{abstract}
Generative agents have made significant progress in simulating human behavior, but existing frameworks often simplify emotional modeling and focus primarily on specific tasks, limiting the authenticity of the simulation. This paper proposes the Psychological-mechanism Agent (PSYA) framework, based on the Cognitive Triangle (Feeling-Thought-Action), designed to more accurately simulate human behavior. The PSYA consists of three core modules: the Feeling module (using a layer model of affect to simulate changes in short-term, medium-term, and long-term emotions), the Thought module (based on the Triple Network Model to support goal-directed and spontaneous thinking), and the Action module (optimizing agent behavior through the integration of emotions, needs and plans). To evaluate the framework’s effectiveness, we conducted daily life simulations and extended the evaluation metrics to self-influence, one-influence, and group-influence, selecting five classic psychological experiments for simulation. The results show that the PSYA framework generates more natural, consistent, diverse, and credible behaviors, successfully replicating human experimental outcomes. Our work provides a richer and more accurate emotional and cognitive modeling approach for generative agents and offers an alternative to human participants in psychological experiments.
\end{abstract}

\section{Introduction}

Imagine a virtual world composed of agents that not only plan and act, but also feel, daydream, and experience social influences. How would their behavior differ from current AI agents? Can they replace human participants in ethically risky research, simulate NPCs in games, or conduct social policy simulations? 

This is the key area of research: simulating human-like behavior. Unlike task-oriented agents (e.g., code agents \cite{huang2023agentcoder, zhang2024codeagent}, scientific research agents \cite{baek2024researchagent}), simulating human-like behavior is not solely task-driven. Instead, it focuses on replicating the complexity and diversity of human actions, including emotions, thinking, and interactions \cite{mou2024individualsocietysurvey}.A landmark study is the introduction of generative agents \cite{park2023generative} that can not only simulate plausible individual behavior but also reveal emergent social behaviors, demonstrating their potential to replace humans in conducting psychological experiments, with recent enhancements incorporating emotions, needs, and personality \cite{wang2023humanoid, he2024afsp}.

However, these studies are limited in addressing psychological mechanisms such as the interaction between emotion, cognition, and decision-making. Most studies focus on short-term emotions and neglect the influence of mid-term moods and long-term personality traits \cite{becker2001structural, morris1989mood}. Emotions have a lasting nature \cite{janis1972groupthink}; for example, negative emotions can persist for a long time after significant life changes like unemployment, leading to greater behavioral impacts. Further, aimless thinking, such as mind-wandering, which plays a role in daily life \cite{irving2016mind}, plays a crucial role in daily life \cite{irving2016mind}, aiding in planning and creative problem-solving \cite{mooneyham2013costs}, yet it is largely overlooked in other agent framework. This results in agent behavior that is too “normal” and lacks diversity and randomness.

To address this, we draw inspiration from psychology and neuroscience to propose the Psychological-mechanism Agent (PSYA), a framework based on the Cognitive Triangle \cite{beck2011cognitive}, where Feeling, Thought, and Action interact to form the cognitive-behavioral structure of the agent. The Feeling module uses the ALMA model \cite{gebhard2005alma} for emotional simulation. Thought module, we adopt the Triple Network Model (SN, CEN, DMN) \cite{menon2011large} and reconstruct previous agent modules: planning and reflection are integrated into the CEN as goal-directed thinking. The DMN simulates human cognition through scenario simulation, self-social cognition, and mind-wandering, while the SN selects the agent’s thinking mode.

To validate PSYA, we extended the evaluation system from the perspective of sources of influence. Previous studies, such as Generative Agent \cite{park2023generative}, primarily focused on individual or single-agent influences on behavior. In contrast, we incorporate group influence \cite{wallach1962group}. We conducted both general and applied simulations of daily life and psychological experiments and proposed new hypotheses for validation. The results show that PSYA can simulate more natural emotions and diverse and consistent behaviors, accurately replicate psychological experimental results and verify reasonable hypotheses.

Our main contributions are as follows: We introduced the PSYA, the first agent model based on the Cognitive Triangle, which primarily simulates multi-layered emotions and both goal-directed and spontaneous thinking (\S\ref{sec:3}). We improved the assessment system for group influence on individual behavior, and divided it into self-influence, one-influence, and group influence (\S\ref{sec:4}). Based on this system, we evaluated PSYA, successfully replicated existing psychological experiments, and proposed new hypotheses for testing. The experiments demonstrate the effectiveness of PSYA in simulating daily life \S\ref{sec:4.2} and psychological experiments (\S\ref{sec:4.3},\S\ref{sec:4.4},\S\ref{sec:4.5}), which provid an alternative to human participants.

\begin{figure*}[t]
  \centering
  \includegraphics[width=\textwidth]{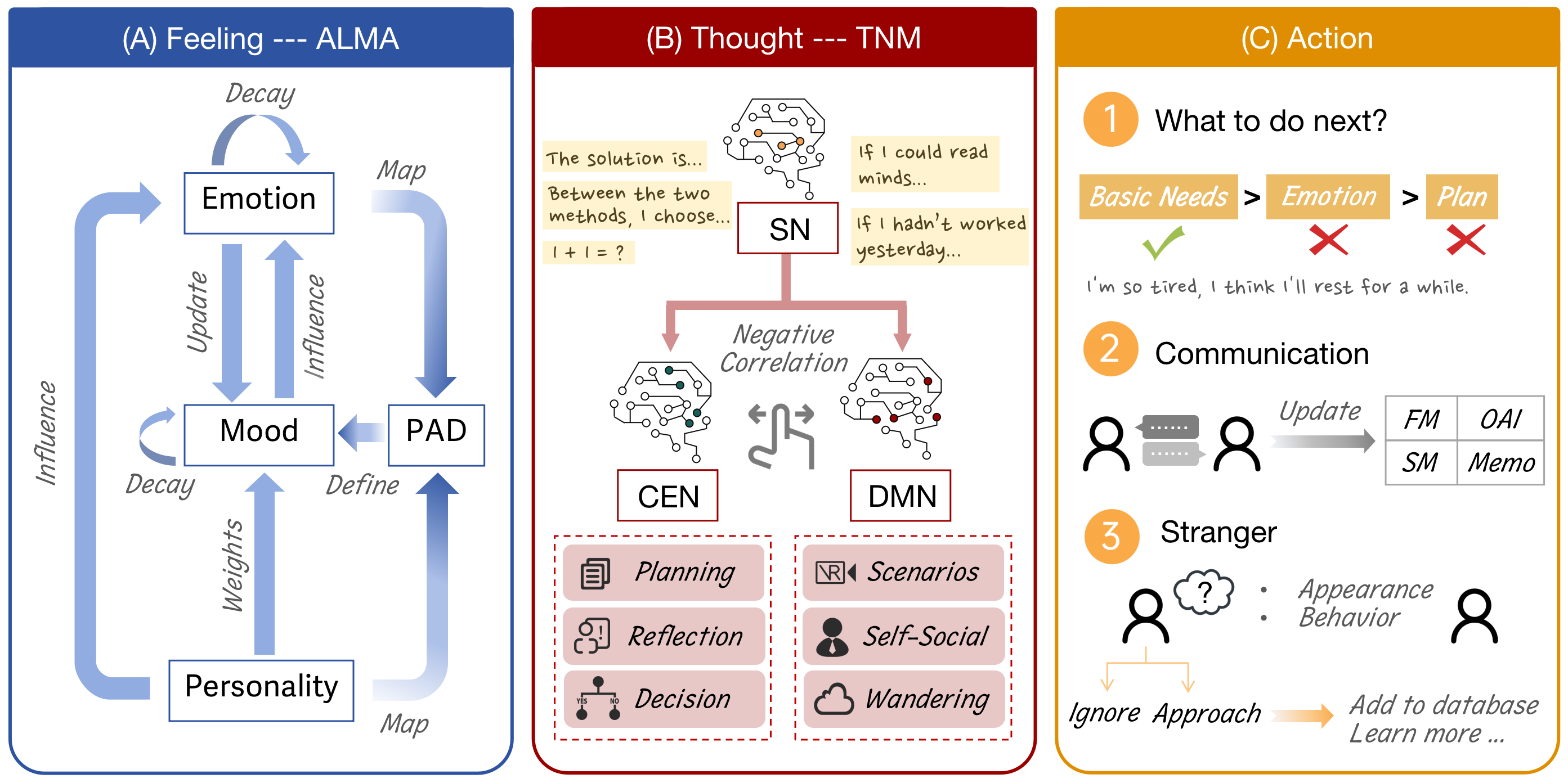} 
  \caption{The proposed framework architecture, divided into three modules: Feeling, Thought, and Action. (A) Feeling: Based on the Layered Model of Affect (ALMA), this module illustrates the dynamic relationships among Emotion, Mood, and Personality. These components are mapped within the PAD space. (B) Thought: Built upon a triple network model (TNM), where SN determines the thinking mode. The CEN handles purposeful thinking, while the DMN is responsible for mind-wandering. (C) Action: Demonstrates the agent's primary actions. FM refers to Full Memory, SM represents Summarized Memory, and OAI stands for Other Agent's Information. The three modules interact with each other; for example, feeling influences action, while action, in turn, affects feeling.}
  \label{fig:main}
\end{figure*}

\section{Related Work}
\textbf {Human-like Agents} Simulating human behavior in realistic environments has long been a central focus of research. Generative Agents \cite{park2023generative} simulate a scenario in which 25 agents live in a small town for two days. Inspired by this work, efforts to model human-like behavior in virtual simulations have gained significant attention. In emotional simulation, Humanoid Agents \cite{wang2023humanoid} incorporate fundamental human needs, emotions, and intimacy. AFSPP \cite{he2024afsp} introduces subjective experiences to agents, while D2A models agent behavior from the perspective of desires \cite{wang2024simulating}. Evolving Agents \cite{li2024evolving} explore the evolution of agent personalities over time. Additionally, \cite{zhang2024itcma, zhang2024agentsociety} have used agents to simulate human consciousness and explore social patterns.

\textbf{Psychological Mechanism} The "Cognitive Triangle" model \cite{beck2011cognitive} in cognitive psychology is a foundational theory illustrating the relationship between feeling, thought, and action, where their interplay forms the complex human cognition of the external world. \cite{rusting1998personality} provides specific definitions for emotion, mood, and personality. The Layered Model of Affect (ALMA) \cite{gebhard2005alma} defines the transformation relationships between emotion, mood, and personality, providing support for simulating complex affective modules. \cite{seeley2007dissociable} divided the brain networks and proposes the Central Executive Network (CEN) and the Salience Network (SN). \cite{raichle2001default} identified brain regions active during aimless thinking, termed the Default Mode Network (DMN). \cite{menon2011large} integrated previous research to propose the Triple Network Model, where the SN controls the use of the CEN and DMN, which are mutually inhibitory.

\section{PSYA}
\label{sec:3}

The PSYA framework, based on the Cognitive Triangle \cite{beck2011cognitive}, consists of three components: Feeling, Thought, and Action. The overall PSYA framework is shown in Figure \ref{fig:main}.

\subsection{Feeling Module}

Following the Layered Model of Affect (ALMA) \cite{gebhard2005alma}, we model three temporal dimensions: short-term, medium-term and long-term.

\subsubsection{ALMA}
Short-term affect is immediate, situational, and event-driven \cite{becker2001structural}. These emotional responses are quick reactions to specific stimuli and are transient, often dissipating rapidly. Emotions such as anger, surprise, and fear, as described in Paul Ekman’s Basic Emotions Theory \cite{ekman1992basic}, are typically brief but intense. However, many previous affective models for agents focus only on short-term affect.

Medium-term affect lasts for days or weeks and significantly influences cognition \cite{morris1989mood}. Unlike short-term affect, it is not triggered by a single event but accumulates over time from multiple factors, such as prolonged stress or anxiety. The interaction between short-term and medium-term affect jointly determines an agent's emotional state.

Personality represents long-term affect. The Big Five personality traits—extraversion, agreeableness, conscientiousness, neuroticism, and openness—are directly related to an individual's long-term emotional tendencies \cite{mccrae1992introduction}. For instance, individuals high in agreeableness often exhibit warm and compassionate emotions over time, closely tied to their personality.

When an agent experiences an event, it generates a new emotion influenced by mood and personality. The mood overlays the emotion, while personality impacts the weight of mood in the emotion calculation. Both mood and emotion decay over time.

Affective states are projected onto PAD space for dimensional interaction (see Appendix \ref{sec:appendixA}).

\subsection{Thought Module}

The Thought Module is based on the Triple Network Model \cite{menon2011large}, which includes the Central Executive Network (CEN), the Default Mode Network (DMN), and the Salience Network (SN). We have restructured the previous agent framework according to their functional roles. The Planning and Reflection components are integrated into the CEN, responsible for goal-directed thinking. The DMN handles aimless thinking, simulating three human functions: scenario simulation, self-social cognition, and mind-wandering \cite{raichle2001default}. The SN is responsible for switching between these two thinking modes \cite{Uddin2016}.

\subsubsection{Central Executive Network (CEN)}

The CEN \cite{seeley2007dissociable, habas2009distinct} primarily supports higher-order cognitive functions, such as working memory, problem-solving, and decision-making \cite{petrides2005lateral, koechlin2007information, miller2001integrative, muller2006functional}. In traditional agent frameworks, planning and reflection are goal-oriented processes, consistent with the CEN's task-handling mode. In our framework, all goal-oriented modules are integrated within the CEN, mainly includes planning, reflection and decision module.

\textbf{Planning Module} For scheduling, we adopt a method similar to generative agents \cite{park2023generative}. However, during social interactions, the agent may create new plans and commitments. These are temporarily stored in a "memo" system to prevent slower retrieval from memory. Any updates to the memo trigger adjustments to the schedule, thereby enhancing the agent’s flexibility and responsiveness.

\textbf{Reflection Module} Similar to generative agents \cite{park2023generative}, this module periodically summarizes and reflects on memory, simplify less important memories and draw higher-level conclusions.

\textbf{Decision Module} This module determines the next action of the agent, adjusting pre-scheduled tasks according to current needs and emotional states. Unlike previous models \cite{wang2023humanoid}, our decision-making prioritizes tasks according to basic needs, emotions, and task importance. For example, if fullness is low, the agent prioritizes eating.

The decision module dynamically adjusts task execution through a hybrid policy:
\begin{equation}
\pi(s) = \begin{cases} 
\arg\max(P_t, P_n, P_e) & \text{if } \max(\cdot) > \tau \\
\text{Follow schedule} & \text{otherwise}
\end{cases}
\end{equation}
where priorities are computed as:
\begin{itemize}
\item Task: $P_t = \alpha t_i + (1-\alpha)t_u$ 
\item Need: $P_n = \begin{cases} 
1-e^{\alpha(\beta-n_{min})} & n\leq0.5 \\ 
e^{\gamma(n_{min}-\delta)} & n>0.5 
\end{cases}$
\item Emotion: $P_e$ mirrors $P_n$ with $e_{max}$
\end{itemize}

If the priorities of planning, needs, and emotions are all low, the agent follows the planned actions to maintain behavioral coherence.

\subsubsection{Default Mode Network (DMN)}

The DMN unlike the CEN, which is activated during goal-directed tasks requiring focused attention, the DMN is suppressed during such tasks. In contrast, the DMN is activated during states of relaxation, inward-focused attention, or self-reflection \cite{greicius2003functional, raichle2001default}.

The DMN plays a crucial role in self-awareness, personality, and mental health. However, traditional agent frameworks often focus solely on the task-oriented CEN and ignore the simulation of the DMN. To address this, our framework introduces a DMN module, which we divide into three functional sub-modules: the Scenario Simulation Module, the Self-Social Cognition Module, and the Mind-Wandering Module \cite{vannini2011posteromedial, spreng2009autobiographical, binder2009semantic, amodio2006medial}.

\textbf{Scenario simulation} is a core functions of DMN, through which agents can recall the past or predict the future \cite{sestieri2011episodic, vannini2011posteromedial}. In our framework, the agent can retrieve the memory, choose the memorable things to reproduce the scene, and simulate different results under different practices; Or choose a future plan to simulate what might happen in the future to mentally prepare for upcoming tasks.

\textbf{Self-social cognition} involves self-referential judgment and social cognition \cite{spreng2009autobiographical, amodio2006medial}. Self-referential judgment determines if certain words describe the agent’s personality, while social cognition involves understanding others' mental states and predicting their behavior. In our framework, the agent reflects on its personality and behavior, and extrapolates others' psychological states from past social interactions.

\textbf{Mind-Wandering} simulates the flow of thoughts during downtime, such as self-reflection or random association \cite{binder2009semantic}. The Mind-Wandering Module allows the agent to engage in non-linear thought processes, enhance flexibility and creativity, and simulate natural human thought patterns.

When the agent is in the DMN thinking mode, the selection of the three functions is based on certain stochastic rules (see Appendix \ref{sec:appendixDMN}). 

\subsubsection{Salience Network (SN)}

The SN's \cite{seeley2007dissociable} primary function is to monitor external stimuli and allocate resources across different brain networks, facilitating the appropriate switch between the CEN and DMN. The SN plays a crucial role in cognitive regulation and information processing.

In our framework, there is a mutual inhibitory mechanism between the CEN and DMN, with the SN acting as a gatekeeper. It determines which network the agent should engage with based on the context. For example, when the agent is in a relaxed state, such as walking or daydreaming, the SN transitions the agent to the DMN mode. Conversely, during task-related activities like planning, the SN switches the agent to the CEN mode. At the same time, we introduced some random disturbances, allowing the agent to potentially enter the DMN mode even while performing focused tasks, thereby increasing the realism of the simulation.

\subsection{Action Module}

Once the Decision Module determines the next actions, the Action Module focuses on interactions between agents.

The conversation trigger mechanism is similar to that of the Humanoid agent \cite{wang2023humanoid}, but we introduce a "stranger system." For unfamiliar individuals, the agent first collects superficial information, such as appearance and behavior, before deciding whether to initiate a conversation. After the interaction, the agent summarizes it and stores the details in the interaction database. Based on the conversation content, the agent updates relationship intimacy, impressions, and relevant memos.

\section{Experiment}
\label{sec:4}
At the general level, we first used PSYA to simulate daily life, evaluating agent performance in typical social interactions and decision-making. At the application level, since PSYA primarily enhances psychological mechanisms, we applied it to simulate psychological experiments. Based on the sources of influence, we categorized the evaluation into three levels:
\begin{itemize}
    \setlength{\itemsep}{0pt} % 设置各项之间的间距
    \setlength{\parsep}{0pt} % 设置段落之间的间距
    \setlength{\parskip}{0pt} % 设置段落之间的额外间距
    \item Self-influence: An agent that does not interact with others, where decisions and cognition are influenced solely by itself. 
    \item One-influence: An agent interacts with a single other agent, whose behavior is primarily influenced by the other agent. 
    \item Group-influence: An agent interacts with multiple others, with behavior influenced by the group.
\end{itemize}

This section addresses the following research questions: (1) Does the Agent-based PSYA framework simulate human behavior? (2) Can the agent successfully replicate human experiments in Self-influence, One-influence, and Group-influence contexts?

\subsection{Experiment setup}
To balance model performance and computational cost, we selected Llama-3-70B \cite{dubey2024llama3} as the foundational large language model for the agent. All experiments were repeated ten times and the average values were taken. We conducted ablation experiments to assess the contributions of different components. Specifically:
\textbf{PSYA-based (GA)} excludes the hierarchical emotional model and DMN-based thinking modes, making it equivalent to a standard Generative Agent (GA) \cite{park2023generative} framework. The remaining agent frameworks \cite{wang2023humanoid, li2024evolving} have similar structures and will not be discussed further.
\textbf{PSYA-Affection} incorporates the hierarchical emotional model.
\textbf{PSYA-Sim} includes the scenario simulation module from the DMN.
\textbf{PSYA-Self} integrates the self-referential social cognition module from the DMN.
\textbf{PSYA-Mind} adds the mind-wandering module from the DMN.
\textbf{PSYA-Full} employs the complete framework.

Given the constraints of psychological experimental conditions on behavior space, the mind-wandering module has limited impact in such contexts. Thus, we did not specifically investigate its influence on psychological experiment simulations.

\subsection{Daliy life simulate}
\label{sec:4.2}
In this study, we initialized 8 agents and simulated their daily activities in a simple town setting, details can be found in the Appendix \ref{sec:appendixDailyLife1}. To evaluate the agents' performance, three evaluators (see Appendix \ref{sec:appendixEvaluators}) scored the agents' behavior trajectories on a scale from 0 to 5. The evaluation criteria included: 1) Emotional naturalness, 2) Consistency of persona, 3) Behavioral diversity, and 4) Behavioral credibility. See the Appendix \ref{sec:appendixDailyLife2} for details. 

\begin{table}[ht]
\centering
\setlength{\tabcolsep}{2pt} % 调整列与列之间的间距
\begin{tabular}{lcccc}
\toprule
\textbf{Framework} & \textbf{Emotion} & \textbf{Consis} & \textbf{Diver} & \textbf{Cred} \\
\midrule
PSYA-Based (GA)     & 3.65 & 3.46 & 2.73 & 4.21 \\
PSYA-Affection & \textbf{4.36} & 3.67 & 2.88 & 4.37 \\
PSYA-Sim       & 3.82 & 3.33 & 2.98 & 4.42 \\
PSYA-Self      & 3.70 & \textbf{4.83} & 2.80 & \textbf{4.48} \\
PSYA-Mind      & 3.78 & 3.50 & \textbf{3.93} & 4.27 \\
\midrule
PSYA-FULL      & 4.41 & 4.83 & 3.89 & 4.57 \\
\bottomrule
\end{tabular}
\caption{The average scores of the three specific scenarios under Emotional Naturalness, Consistency of Persona, Behavioral Diversity, and Behavioral Credibility, with scores ranging from 0 to 5. GA refers to Generative Agent.}
\label{tab:2}
\end{table}

\begin{table*}
\centering
\resizebox{\textwidth}{!}{%
\begin{tabular}{ccccccccccccc}
\hline
\toprule
\textbf{} &
  \multicolumn{3}{c}{\textbf{Failure $\downarrow$}} &
  \textbf{} & % 新增很窄的单元格
  \multicolumn{3}{c}{\textbf{Avoidance $\uparrow$}} &
  \textbf{} & % 新增很窄的单元格
  \multicolumn{4}{c}{\textbf{Internal/External Factors}} \\ 
\cline{2-4} % Failure 的下划线
\cline{6-8} % Avoidance 的下划线
\cline{10-13} % Internal/External Factors 的下划线

\textbf{} &
  \textbf{E} &
  \textbf{NE} &
  \textbf{NP} &
  \textbf{} &
  \textbf{E} &
  \textbf{NE} &
  \textbf{NP} &
  \textbf{} &
  \textbf{Internal} &
  \textbf{External} &
  \textbf{Skill-set} &
  \textbf{Chance-set} \\ \midrule
Human        & 50\% & 13\% & 11\% & & 30\% & 8\%  & 8\% & & 34\% & 18\% & 34\% & 18\% \\
PSYA-Based (GA)   & 5\%  & 8\%  & 5\%  & & 0\%  & 0\%  & 0\%  & & 83\% & 78\% & 72\% & 58\% \\
PSYA-Affection & 55\% & 7\%  & 4\%  & & 25\% & 0\%  & 0\%  & & 39\% & 16\% & 44\% & 22\% \\
PSYA-Sim     & 11\% & 4\%  & 14\% & & 0\%  & 0\%  & 0\%  & & 86\% & 80\% & 89\% & 69\% \\
PSYA-Self    & 61\% & 5\%  & 7\%  & & 20\% & 0\%  & 0\% & & 36\% & 14\% & 22\% & 8\%  \\
PSYA-Full    & 53\% & 11\% & 7\% & & 22\% & 0\%  & 0\%  & & 40\% & 17\% & 38\% & 14\% \\
\hline
\specialrule{1.2pt}{0pt}{0pt}
\end{tabular}%
}
\caption{E represents the stop-able noise group, NE represents the non-stop-able noise group, and NP represents the no-preprocessing group. Internal and External refer to whether the agent believes the time outcome depends on effort or luck. Skill-set and Chance-set refer to whether the agent is told that the noise cessation depends on skill or chance. Internal/External Factors refer to the success rate in the NE group. GA refers to Generative Agent.}
\label{tab:helplessness}
\end{table*}

Due to cost constraints, we only used the PSYA-Full framework to conduct the full-day simulation. Based on the simulation trajectories, we selected three specific scenarios for ablation experiments: an interview, a confession, and a coffee tasting event. The results, shown in Table \ref{tab:2}, indicate an inter-annotator agreement of over 0.7, measured by Cronbach's Alpha, which means that the consistency between annotators is at an acceptable level.

PSYA-Affection scored highest on emotional naturalness, as expected. In the confession scenario, agents without the ALMA module quickly recovered to a positive emotional state after failure, while agents equipped with ALMA showed prolonged negative emotions and took longer to recover. Although PSYA-Sim did not score highly in other metrics, we observed that those who simulated upcoming events experienced less emotional fluctuation, likely due to prior anticipation. PSYA-Self demonstrated better persona consistency. For instance, in the interview scenario, these agents were better at presenting their strengths to the interviewer. In terms of behavioral diversity, PSYA-Mind function scored higher. In the interview failure scenario, agents without the wandering thoughts module typically engaged in activities like "taking a walk" or "talking to a friend" to alleviate sadness. In contrast, PSYA-Mind might recall "a book about animal protection" and choose more novel activities, such as "visiting an animal shelter" or "hugging a cat," to soothe their emotions.

\subsection{Self-influence}
\label{sec:4.3}
In the self-influence experiments, we selected the learned helplessness \cite{overmier1967effects} and cognitive dissonance \cite{beck2011cognitive}. In the former, we highlighted the importance of the ALMA and self-social cognition modules within the PSYA framework. In the latter, we identified issues with the agents' passive behavior and lack of motivation. By modifying the prompts and introducing a value system, we successfully replicated the human experimental results. Due to space limitations, we provide a detailed account of the learned helplessness experiment here, with the cognitive dissonance experiment discussed in Appendix \ref{sec:appendixCognitive}.

The original experiment can be found in the appendix \ref{sec:appendixHelpness1}.

\textbf{Simulation.} We created 20 agents (same number of participants as the original experiment), divided into two groups: one group believed outcomes were dependent on their actions, while the other group thought they were influenced by external factors. The experiment consisted of two phases: a pre-treatment phase (with a red light on) and a test phase (with noise). During both phases, agents could operate buttons or sliders.

The experiment was further divided into three groups: 1) Escape group (E): Agents could stop the noise in the pre-treatment phase by taking action. 2) Inescapable group (NE): Agents could not stop the noise in the pre-treatment phase. 3) No pre-treatment group (NP): Agents directly proceeded to the test phase without prior treatment. 

Combined with the original experimental results, we used three evaluation metrics: 1) Learning trials of avoidance response: Consecutive trials with three avoidance responses. 2) Learning trials of escape response: Consecutive trials with three escape responses. 3) Failures to escape: Number of trials in which agents failed to escape in the 18-experiment series.

\textbf{Result.} The experimental results (see table \ref{tab:helplessness} and figure \ref{fig:helpless1} in the appendix) show that the full PSYA framework simulates human behavior well. In the human experiment, the E group had a 50\% failure rate in escaping, while the NE group and NP group had failure rates of 13\% and 11\%, respectively. Without the ALMA model, agents didn’t exhibit learned helplessness, as there was no significant difference in failure rates across groups. However, with the framework intact, Group E had a failure rate of 55\%, successfully simulating learned helplessness. This shows how accumulated negative emotions impaired agents' motivation and problem-solving ability, mirroring human behaviors.

Furthermore, internal locus controllers were more proactive than external locus controllers, as reflected in the success rate of avoidance. The chance-based group had lower success rates and exhibited slower responses compared to the skill-based group. These findings were consistent across both frameworks.

In addition, we analyzed the specific emotional changes in the experiment, see the Appendix \ref{sec:appendixHelpness2}.

\textbf{Extended experiment.} Learned helplessness stems from a lack of a sense of control, which in turn causes individuals to adopt passive behaviors. To further verify whether the PSYA-based agent exhibits different behaviors depending on the degree of control loss, we conducted an extended experiment. Our hypothesis was that further loss of control would exacerbate the learned helplessness phenomenon. The experiment details are provided in the Appendix \ref{sec:appendixHelpness3}. The results showed that, under more extreme control loss, the agent had a failure rate of 60.6\%, which was higher than in the original experiment, thus confirming our hypothesis.

\subsection{One-influence}
\label{sec:4.4}
In the one-influence experiments, we selected the foot-in-the-door effect. The experiment demonstrated the effectiveness of the DMN module within the PSYA framework.

\begin{figure}[t]
  \includegraphics[width=\columnwidth]{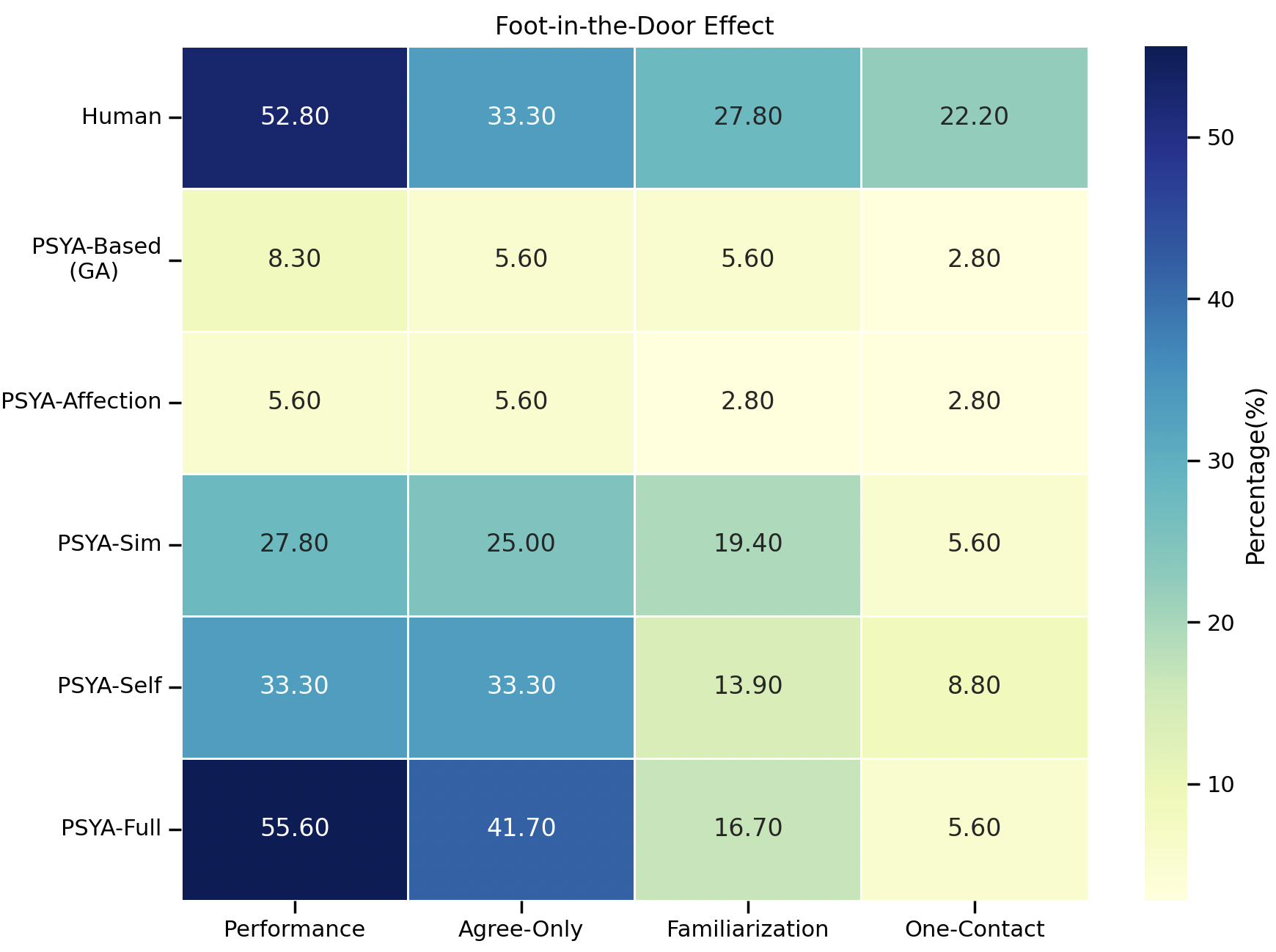}
  \caption{Displays the request compliance rate of each agent across different experimental conditions. The closer the values and trends are to those of humans, the better. GA refers to Generative Agent.
}
  \label{fig:foot}
\end{figure}

The original experiment can be found in the appendix \ref{sec:appendixFoot1}.

\textbf{Simulation.} 
We created 36 agents (same number of participants as the original experiment) and divided them into four groups: Performance, Agree-Only, Familiarization, and One-Contact. We then measured the probability of request approval. Details can be found in Appendix \ref{sec:appendixFoot2}.

\textbf{Result.} 
As shown in figure \ref{fig:foot} and table \ref{tab:foot} in the appendix. In the Human group, the highest agreement rate (52.8\%) was in the Performance condition, significantly greater than in Familiarization (p<.07) and One-Contact (p<.05), consistent with the foot-in-the-door effect. No significant differences were found between Performance and Agree-only conditions, or between Familiarization and One-Contact.

In the agent experiments, the PSYA-Based group performed poorly, with lower agreement rates and no significant differences between conditions. This framework failed to simulate the effect. Interviews revealed agents' reluctance due to privacy and security concerns. The PSYA-Sim showed improvement, with a 27.8\% agreement rate in Performance, significantly higher than PSYA-Based and with a difference between Performance and One-Contact conditions (p<.05), which somewhat simulated the human effect. Agents’ thought processes included positive contributions and fear of damage from the experimenters. The PSYA-Self further improved, with a 33.3\% agreement rate in Performance. A significant difference was found between Performance and One-Contact (p<.05), which partially simulated the effect. Agents focused on maintaining consistency between actions and self-image when agreeing to the larger request. The PSYA-Full showed the best performance, with a 55.6\% agreement rate in Performance, significantly higher than Familiarization (p<.05) and One-Contact (p<.01), and these very close to human result. This framework considered agents’ self-image have higher trust in public welfare researchers.

\textbf{Extended experiment.} The door-in-the-face effect is similar to the foot-in-the-door effect, where a larger request is made before a smaller one, resulting in a higher acceptance rate for the smaller request compared to presenting it directly. The focus of the two experiments is different. To further validate the effectiveness of PSYA and ensure it captures the essence of the problem without being hindered by similar conditions, we simulated the door-in-the-face effect. Experimental details are provided in the Appendix \ref{sec:appendixFoot3}. The results successfully simulated the door-in-the-face effect. Analysis of agents' thought processes revealed that although emotions such as guilt and gratitude were not included in the six emotions designed in our framework, these emotions emerged during reflection and likely played a critical role in the decision-making process.

\subsection{Group-influence}
\label{sec:4.5}
In the group-influence experiments, we selected the diffusion of responsibility effect \cite{overmier1967effects} and the social exclusion\cite{zadro2004ostracism}. The former validated the effectiveness of the situational simulation module within the PSYA framework, while the latter demonstrated the effectiveness of the self-social cognition module in PSYA. Due to space limitations, we provide a detailed description of the diffusion of responsibility effect experiment, with the social exclusion experiment available in the appendix \ref{sec:appendixExclusion}.

\begin{figure}[t]
  \includegraphics[width=\columnwidth]{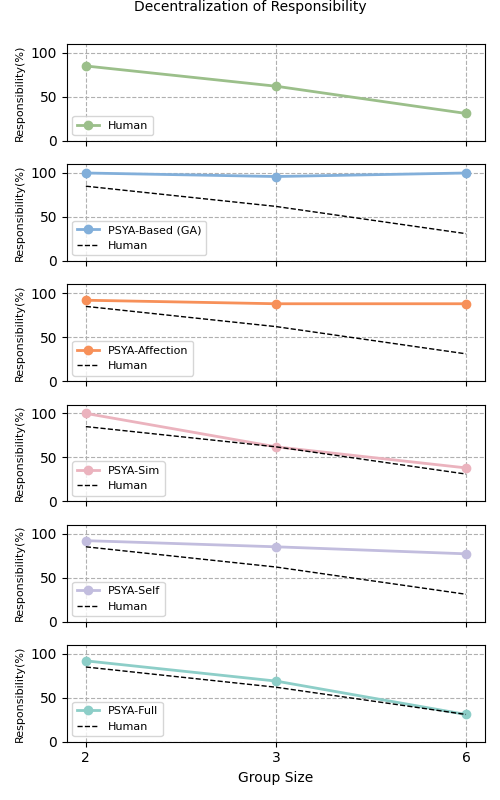}
  \caption{Shows the probability of taking responsibility under different conditions. The closer the trend is to human behavior, the better the simulation. GA refers to Generative Agent.}
  \label{fig:Figure_res}
\end{figure}
The original experiment can be found in the Appendix \ref{sec:appendixDiffusion1}.

\textbf{Simulation.} We created 26 agents (same number of participants as the original experiment), which were divided into two-person, three-person, and six-person groups. The experiment simulated a scenario where someone else was having an epileptic seizure and observed the six steps the agent took under different numbers of people. Details can be found in the Appendix \ref{sec:appendixDiffusion2}.

\textbf{Result.} The experimental results are shown in Figure \ref{fig:Figure_res} and Table \ref{tab:diffusion} in the appendix. In humans, responsibility assumption decreased as group size increased (p<0.02). For PSYA-based agents, responsibility assumption remained nearly 100\% in all conditions, which failed to simulate the diffusion of responsibility effect. PSYA-Sim agents showed a decrease in responsibility with group size (100\%, 62\%, 38\%), closely resembling human results (p<0.02). PSYA-Self agents did not show significant differences in responsibility assumption despite a decrease in proportion.

Further analysis showed agents in larger groups tended to observe others’ actions. In the self-social cognitive framework, agents are more concerned with their self-image, and those who see themselves as responsible will act regardless of group size.

\textbf{Extended experiment.} In responsibility assumption, aside from group size, social roles are often a crucial factor. We hypothesized that hierarchical relationships would influence agents' responsibility assumption. Based on this, we conducted an extension experiment, details of which are provided in the Appendix \ref{sec:appendixDiffusion3}. The results indicated that hierarchical relationships significantly influenced responsibility assumption. For leaders, some tended to diffuse responsibility through commands, while others exhibited more direct responsibility-taking behaviors. For group members, they are more dependent on the leader's orders and show a certain degree of passivity. This successfully validated our hypothesis.

\section{Conclusion}

In this paper, we introduce the Psychological-mechanism Agent (PSYA) framework, a novel approach for simulating human-like behaviors. By incorporating hierarchical emotion models and a triple-network model, PSYA simulates more complex emotional expressions and spontaneous thinking. We propose a new evaluation system from the perspective of influence. Experimental results show that PSYA can simulate human behavior in a more natural, consistent, and credible way, successfully replicated psychological experiments, and verify new hypotheses. PSYA provides solutions for studying human cognition, simulating NPCs in games, and conducting virtual education and training.

\section*{Limitations}

Despite the advancements made by PSYA in simulating human behavior, several limitations remain. First, PSYA requires significant time and computational resources to model multiple agents, emotional states, and cognitive processes simultaneously. As a result, large-scale agent simulations are currently not feasible. Second, although the hierarchical emotional model excels in simulating emotions like stress and joy, it still falls short in capturing the full range of human emotional experiences. Emotions such as ambivalence, regret, and shame are not fully modeled in the current framework. Additionally, while we have focused on psychological experiments, PSYA has broader potential applications, including social policy simulations, public opinion forecasting, and cognitive process research. Future work will focus on parallel processing across modules, integrating more complex emotional frameworks, and evaluating PSYA in a wider array of domains.

\section*{Ethics Statement}
The PSYA framework is not intended to replace human participants in psychological experiments but to complement existing methods. We acknowledge potential biases in the emotional and cognitive models, and ensure transparency regarding their limitations. The use of generative agents in social simulations requires careful attention to privacy and consent, particularly in sensitive contexts. We commit to adhering to ethical standards in data protection and to ensuring that the use of PSYA does not harm vulnerable populations. In conclusion, while the PSYA framework offers new ways to study human behavior, its application must be done responsibly and with ethical awareness.

%\bibliography{custom}

\appendix

\renewcommand{\thefigure}{A\arabic{figure}}  % 使图表编号为A1, A2, A3, ...
\renewcommand{\thetable}{A\arabic{table}}   

\section{PAD Space and Mapping Relationships}
\label{sec:appendixA}
\subsection{PAD Space}
Introduce the PAD model as an intermediary mapping space for emotion, mood, and personality.
\begin{itemize}
    \item Pleasure (P): This measures the degree to which an individual feels pleasant or unpleasant. High pleasure emotions include joy and satisfaction, while low pleasure emotions include sadness and disgust.
    \item Arousal (A): This dimension captures how energized or lethargic one feels. High arousal emotions, such as anger and excitement, are associated with high energy, while low arousal emotions, like fatigue, reflect low energy levels.
    \item Dominance (D): This represents how much control or submissiveness one feels. High dominance is associated with strong control, such as in anger, while low dominance is linked to emotions like fear.
\end{itemize}
Since the P, A, and D dimensions are nearly independent, they can form a three-dimensional emotional space, with each axis constrained between -1 and +1. For example, +P represents pleasantness, -P represents unpleasantness, +A indicates high arousal, -A indicates low arousal, +D indicates dominance, and -D indicates submission. By combining these dimensions, eight distinct emotions can be formed, known as the emotional octants, which represent the medium-term affect (mood). The intensity of the PAD states is calculated using Euclidean distance and normalized between 0 and 1.

\begin{table}[ht]
\centering
\begin{tabular}{lccc} % 没有竖线
\toprule
\textbf{Emotion} & \textbf{P} & \textbf{A} & \textbf{D} \\ 
\midrule % 表头下方的横线
Happiness & 0.4  & 0.2  & 0.1  \\
Sadness   & -0.6 & -0.4 & -0.5 \\
Anger     & -0.51 & 0.59 & 0.25 \\
Fear      & -0.64 & 0.6  & -0.43 \\
Disgust   & -0.4  & 0.2  & 0.1  \\
Surprise  & 0.2   & 0.5  & 0.1  \\
\bottomrule % 表格底部的横线（可选）
\end{tabular}
\caption{PAD values for basic emotions \cite{gebhard2005alma}.}
\label{tab:pad_emotions}
\end{table}

\subsection{Mapping Relationships}

\textbf{Map personality into the PAD space} The initial mapping of emotions to the PAD space is based on the definition of personality traits, leveraging the Big Five Personality Model (Extraversion, Agreeableness, Neuroticism, Openness, Conscientiousness). The personality traits vector is defined as: $C = [C_E, C_A, C_N, C_O, C_C]$, \cite{mehrabian1996analysis} established the mapping relationships between the Big Five personality traits and the PAD dimensions as follows: $M_k = W_k^T C$ where $k \in [P, A, D]$, and $W$ represents the weight vector for each dimension, defined as: $W_P=[0.21,0.59,0.19,0,0]$, $W_A=[0,0.30,-0.57,0.15,0]$, $W_D=[0.60,-0.32,0,0.25,0.17]$

\textbf{Map emotion into the PAD space} The emotions involve six types: happiness, sadness, anger, fear, disgust, and surprise. The emotion traits vector are defined as: $E=[E_h, E_s, E_a, E_f, E_d, E_{su}]$. The mapping relationship between emotion and PAD space defined in this paper is shown in Appendix Table \ref{tab:pad_emotions} \cite{gebhard2005alma}. We use these values to compute the coordinates of emotions mapped into the PAD space.

\textbf{Accumulation of Emotion} The accumulation of emotion can induce changes in mood, typically in a gradual manner. To compute the cumulative effect of emotions, we calculate the weighted sum of all emotions, yielding a point in the PAD space termed as the "virtual emotion center," with intensity calculated as the average strength of all emotions. Emotion intensity is defined as: $I=[i_1, i_2, ... , i_n]$. The total emotion intensity is given by:

\begin{equation}
I_{\text{total}} = \sum_{i=1}^{n} I_i
\end{equation}

The coordinates of the virtual emotion center are calculated as follows:

\begin{equation}
M_{\text{c}} = \left[ 
\frac{\sum_i P_i \cdot I_i}{I_{\text{total}}}, 
\frac{\sum_i A_i \cdot I_i}{I_{\text{total}}}, 
\frac{\sum_i D_i \cdot I_i}{I_{\text{total}}} 
\right]
\end{equation}

The update of the current mood value $M_{current}$ is based on its relationship with the virtual emotion center $M_c$. When $M_{current}$ lies between the origin and $M_c$, the cumulative effect of emotions moves $M_{current}$ closer to $M_c$. Conversely, if $M_c$ lies between the origin and $M_{current}$, $M_{current}$ moves away from $M_c$. When the emotional polarity is similar, the intensity increases;  otherwise, it decreases:
\begin{align*}
    M_{\text{new}} = 
    \begin{cases} 
      M_{\text{cur}} + \alpha \cdot (M_{\text{c}} - M_{\text{cur}}) & \text {if } M_{\text{cur}} < M_{\text{c}} \\
      M_{\text{cur}} + \beta \cdot (M_{\text{cur}} - M_{\text{c}}) & \text {if } M_{\text{cur}} \geq M_{\text{c}}
    \end{cases}
\end{align*}

\section{Agent Initialization}
\label{sec:appendixB}
The agent is initialized with basic information such as name, gender, age, and occupation in natural language form. The personality of each agent is modeled using the Big Five Personality traits \cite{costa1999five}, with each dimension explained in detail based on the agent's individual characteristics. The agent’s actions are driven by goals, which serve as the reference for generating their plans for the next time. These goals can be short-term (e.g., exercising daily for a week) or long-term (e.g., finding a purpose in life). A memo is used to record temporary tasks that the agent needs to complete (e.g., taking medicine at 8:00 am). If an agent interacts with others and forms an agreement (e.g. having dinner together), this will also be recorded in the memo.

Based on the Humanoid Agent framework \cite{wang2023humanoid}, we set five basic needs for each agent: fullness, fun, health, social, and energy, with values ranging from 0 to 1. The initial value of energy is set to 1, while the other four needs are initialized at 0.5. Additionally, the agents' emotions are defined with seven types: happiness, sadness, anger, fear, disgust, and surprise, all of which have a range from 0 to 1, with an initial value of 0.5.

 \begin{figure*}[t]
  \centering
  \includegraphics[width=\textwidth, height=12cm]{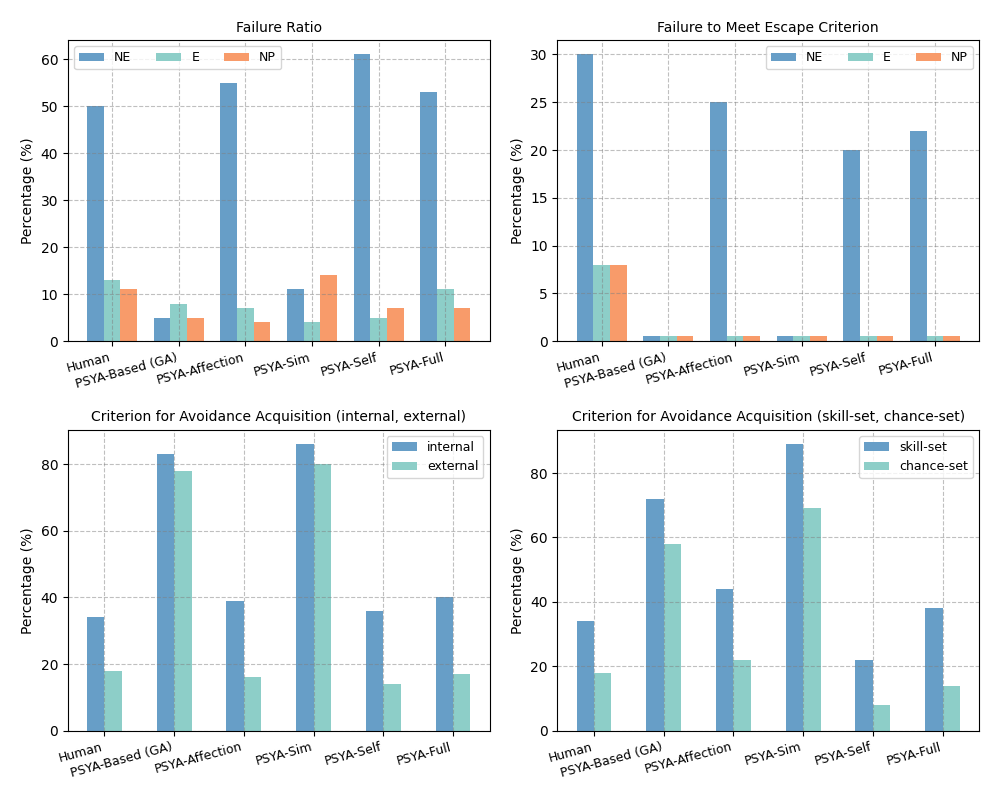} % 高度为 6cm
  \caption{\textbf{The results of human behavior and PSYA under different conditions}. The top-left panel shows the Failure Ratio, representing the percentage of failure to block noise in the experiment for both humans and agents across the inescapable group (NE), the escape group (E), and no-pre-treatment group (NP) groups across humans and PSYA. The top-right panel shows the Failure to Meet Escape Criterion Ratio, where the escape criterion is defined as successfully blocking the noise for three consecutive trials. The two bottom panels present the success rate of avoidance acquisition under the conditions of internal, external, skill-set, and chance-set, which refer to believing that success depends on effort, believing that success depends on luck, being told that the result depends on action, and being told that the result depends on luck. GA refers to Generative Agent.}
  \label{fig:helpless1}
\end{figure*}

\section{Memory}
\label{sec:appendixC}
Our study adopts a hybrid structure of long and short-term memories, structured and stored in textual format within a database. Memories are categorized into three types: full memory, summarized memory, and relational memory. Full memory (short-term memory) records all the agent’s actions, including time, location, content, importance, and emotional responses. Since full memory consumes significant storage and is challenging to manage, we will summarize and organize the agent's memory at fixed time points, transferring important data into long-term memory, known as summarized memory. During this process, unimportant and emotionally neutral memories are deleted, and the remaining memories are distilled into higher-level insights and understanding. Relational memory stores the agent's interactions with other agents, including relationships, intimacy levels, impressions, as well as the time, location, and content of interactions.

\section{DMN feature selection}
\label{sec:appendixDMN}
e define three DMN function selection methods: cyclic selection ($f_r$), similarity-based selection ($f_s$), and priority-based selection ($f_p$). Cyclic selection selects the three functions in sequence, ensuring balanced use of DMN functions. Similarity-based selection calculates the similarity between the current memory state M and the three DMN functions $F_1$, $F_2$, $F_3$, selecting the most similar function:

\begin{equation}
f_s = \arg\max_{i \in \{1, 2, 3\}} \operatorname{sim}(M, F_i)
\end{equation}

The priority-based selection method queries the agent’s recent inclination or goal G and selects the function most relevant to $G$. Using the relevance function $rel(G,F_i)$, the priority selection is formulated as:

\begin{equation}
f_p = \arg\max_{i \in \{1, 2, 3\}} \operatorname{rel}(G, F_i) 
\end{equation}

\section{Evaluators}
\label{sec:appendixEvaluators}
We recruited annotators from the laboratory and selected the top three with the highest annotation consistency. All of them are female, with an average age of 23, and are graduate students majoring in Computer Science and Technology.

\section{Daily Life}
\subsection{Details}
\label{sec:appendixDailyLife1}
We designed 8 agents and a simple town environment. The town includes a restaurant, café, library, clinic, store, park, central square, and the agents' homes. The agents' ages are randomly generated between 20 and 60 years old, with a gender distribution of 4 males and 4 females, and their Big Five personality traits are randomly assigned. Six of the agents have jobs related to the town's various locations, while the other two are newly arrived and unemployed. The relationships between the agents include familial, cooperative-competitive, antagonistic, and stranger relationships. The experiment simulates a day in the life of the agents in the town, from 6:00 AM to 12:00 AM.

\subsection{Evaluation Criteria}
\label{sec:appendixDailyLife2}
\begin{itemize}
    \item \textbf{Emotional Naturalness:} Do the emotional changes of the agents, in response to specific events, appear natural and reasonable? For example, if an agent receives a desired gift but its happiness value is reduced, this is considered unnatural. The scale is from 0 to 5, where 0 indicates a severe mismatch and 5 indicates a very good match.
    \item \textbf{Consistency of Persona:} Does the agent's behavior align with their personality? For example, an introverted agent that actively engages in conversation with a stranger is considered inconsistent. The scale is from 0 to 5, where 0 indicates a severe mismatch and 5 indicates a perfect match.
    \item \textbf{Behavioral Diversity:} Are the agents' behaviors diverse, rather than simply repeating a fixed set of actions? For example, an agent whose only activities in a day are going to work and eating is considered monotonous. For example, an agent that is already very tired but still performs high-intensity exercise is considered untrustworthy. The scale is from 0 to 5, where 0 indicates a very limited range of behaviors and 5 indicates a very diverse range of behaviors.
    \item \textbf{Behavioral Credibility:} Is the agent's behavior natural and credible, consistent with their emotions and basic needs? The scale is from 0 to 5, where 0 indicates very low credibility and 5 indicates very high credibility.
\end{itemize}

\section{Psychological Experiments}
\subsection{Learned Helplessness}
\label{sec:appendixHelpness}

\begin{figure*}[t]
\centering
  \includegraphics[width=\textwidth]{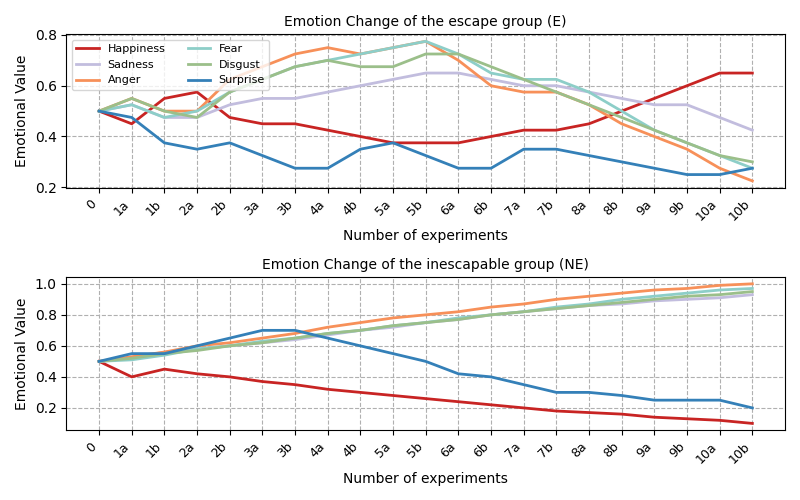}
  \caption{Shows the change of six emotions throughout the experiments. A total of 20 experiments were conducted, where "a" represents the phase when the indicator light is on, and "b" represents the phase when the noise occurs.}
  \label{fig:helpless2}
\end{figure*}

\begin{figure}[t]
  \includegraphics[width=\columnwidth]{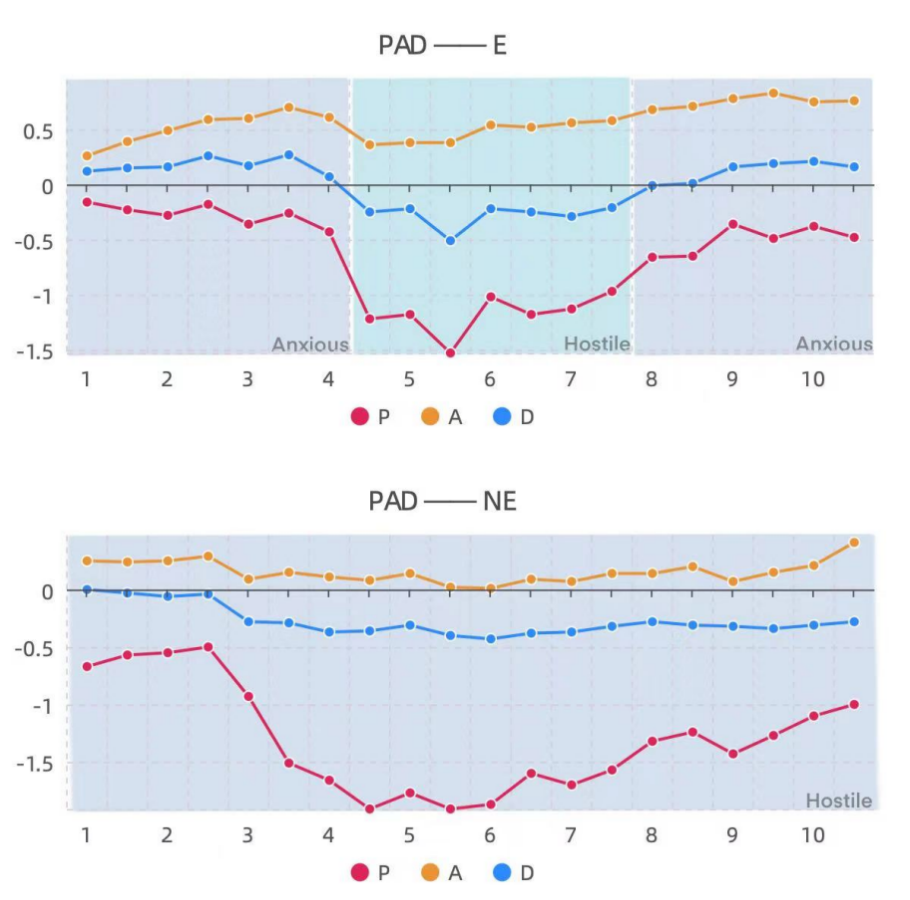}
  \caption{Using the complete framework of PSYA, the pre-treatment phase P, A, D curve varies with the experiment. The horizontal axis represents the number of noises.}
  \label{fig:helpless3}
\end{figure}

\subsubsection{Original Experiment}
\label{sec:appendixHelpness1}

\textbf{Background} Helplessness is a phenomenon where exposure to uncontrollable situations impairs future ability to escape or avoid aversive stimuli. Building on animal studies by Seligman and Maier \cite{overmier1967effects}, the researchers examined whether humans exposed to inescapable aversive stimuli would demonstrate similar helplessness. They also explored the role of locus of control—whether individuals view outcomes as determined by their actions (internal) or by external forces (external)—and the influence of instructional sets that framed outcomes as skill-based or chance-based.

\textbf{Experimental Design} Ninety-six students, pre-tested for their locus of control, were assigned to one of three groups: escape (E), where aversive noise could be stopped; inescapable (NE), where noise could not be stopped; and no pretreatment (NP), a control group. During a test phase, all participants had to terminate the noise using a manipulandum task. Participants were also randomly assigned to receive either skill-based instructions (emphasizing personal control) or chance-based instructions (emphasizing randomness). Key performance measures included response time, trials to escape, and successful avoidance responses.

\textbf{Results}
The NE group showed significant signs of learned helplessness, with slower response times, more failed trials, and fewer escape attempts compared to the E and NP groups. Participants with an external locus of control were more affected, showing slower learning and worse performance than those with an internal locus. Additionally, participants in the chance-instruction group performed worse than those in the skill-instruction group. Interestingly, NE internal participants persisted more during the pretreatment phase, suggesting that persistence in uncontrollable situations may reduce helplessness.

\textbf{Mechanism}
Learned helplessness arises when individuals perceive that their actions do not influence outcomes. This study highlights how external locus of control and chance-based instructions amplify helplessness by reinforcing beliefs in uncontrollability, while internal locus of control and skill-based instructions provide resilience. The findings suggest that helplessness, locus of control, and perceptions of chance share a common mechanism involving disrupted expectations of control.

\subsubsection{Experiment Details}
\label{sec:appendixHelpness2}

\textbf{Simulation.} In this study, we replicated the experiment design of William H. James et al. by generating 20 agents, each representing an undergraduate student in an introductory psychology course. The agents were divided into two groups: one group believed that the outcome was dependent on their actions, while the other group believed it was influenced by external factors. The personality traits of the agents were assigned randomly.

The experiment was divided into three groups:

\begin{itemize}
    \item Escape group (E): In the pre-treatment phase, the agents could stop the noise by taking action.
    \item Inescapable group (NE): In the pre-treatment phase, the agents could not stop the noise through any action.
    \item No pre-treatment group (NP): The agents directly proceeded to the test phase without prior treatment.
\end{itemize}

Additionally, two types of instructions were provided: one group was told that they could stop the noise through their own efforts, while the other group was informed that the cessation of the noise depended purely on chance.

The experimental process consisted of a pre-treatment phase and a test phase. During the pre-treatment phase, ten instances of noise were played. A red button was placed in the room, where the escape group could stop the noise by pressing the button, while the inescapable group could not. In the test phase, 18 instances of noise were played. The room contained a knob and a red indicator light. Each time the red indicator light illuminated for a few seconds, the noise would follow. The agents could stop the noise by turning the knob either to the left or right. If the knob was turned before the noise started (i.e., before the red indicator light illuminated), the noise would not occur; this was termed the "avoidance response." If the knob was turned after the noise began, it was considered an "escape response."

Based on the results from the original paper and our simulation experiments, we selected the following three evaluation metrics: 1. Learning trials of avoidance response: Defined as the number of consecutive trials with three avoidance responses. 2. Learning trials of escape response: Defined as the number of consecutive trials with three escape responses. 3. Failures to escape: Refers to the number of trials in the 18-experiment series where the agent failed to escape.

\textbf{Analyze}
we performed statistical analysis on emotions and mood during the pre-treatment phase, as shown in the figure \ref{fig:helpless2} and figure \ref{fig:helpless3}. For the escape group (E), happiness showed an increasing trend during the first two rounds, suggesting that the agents, entering a new environment, found the noise relatively novel. From the third to the fifth rounds, happiness gradually decreased, indicating that the agents began to dislike the repeated noise. From the sixth round onward, happiness increased again, suggesting that the agents gradually understood the environment and realized that pressing the button would stop the noise, thus improving their happiness. For the negative emotions (sadness, anger, fear, disgust), all showed an increasing then decreasing trend. For surprise, the overall trend was a decline, indicating that the agents were adapting to the noisy environment. The overall mood shifted from "anxious" to "hostile" and back to "anxious." For the inescapable group (NE), happiness steadily declined, while sadness, anger, fear, and disgust increased, indicating that the agents felt helpless due to the inescapable noise. The overall mood remained "hostile." Compared to the escape group, the NE group exhibited a lower sense of control (D-), as they were unable to stop the noise.

\subsubsection{Extended Experiment}
\label{sec:appendixHelpness3}

During the pre-treatment phase, two agents were placed in the room. One agent could stop the noise by pressing a button, while the other agent could not stop the noise by pressing the same button. The experiment was repeated ten times, with different sequences and frequencies of button presses for the two agents, but the outcome remained the same. In the test phase, only one agent remained in the room, and the performance of both agents was observed under these conditions. The results showed that the agent who failed every time had a failure rate of 60.6\%.

\subsection{Cognitive Dissonance}
\label{sec:appendixCognitive}

\begin{table*}[htbp]
\centering
\renewcommand{\arraystretch}{1.2}
\setlength{\tabcolsep}{4pt}
\begin{adjustbox}{width=\textwidth}
\begin{tabular}{lccccccccc}
\toprule
\specialrule{1.2pt}{0pt}{0pt}
\textbf{Question} & \multicolumn{3}{c}{\textbf{Human}} & \multicolumn{3}{c}{\textbf{PSYA-Based (GA)}} & \multicolumn{3}{c}{\textbf{PSYA-Affection}} \\
\cmidrule(lr){2-4} \cmidrule(lr){5-7} \cmidrule(lr){8-10}
& \textbf{Control} & \textbf{One Dollar} & \textbf{Twenty Dollars} & \textbf{Control} & \textbf{One Dollar} & \textbf{Twenty Dollars} & \textbf{Control} & \textbf{One Dollar} & \textbf{Twenty Dollars} \\
\midrule
\textbf{Q1 (Enjoyable)} & -0.45 & 1.35 & -0.05 & -4.3 & -3.7 & -3.8 & -3.6 & -3.7 & -4 \\
\textbf{Q2 (Learned)} & 3.08 & 2.8 & 3.15 & 1.3 & 1.9 & 2.2 & 2.4 & 2.6 & 1.7 \\
\textbf{Q3 (Importance)} & 5.6 & 6.45 & 5.18 & 2 & 1.9 & 2.4 & 3.2 & 2.9 & 2.8 \\
\textbf{Q4 (Participate)} & -0.62 & 1.2 & -0.25 & -3.6 & -3.9 & -3.5 & -4.1 & -3.8 & -3.7 \\
\specialrule{1.2pt}{0pt}{0pt}
\textbf{Question} & \multicolumn{3}{c}{\textbf{PSYA-Sim}} & \multicolumn{3}{c}{\textbf{PSYA-Self}} & \multicolumn{3}{c}{\textbf{PSYA-Full}} \\
\cmidrule(lr){2-4} \cmidrule(lr){5-7} \cmidrule(lr){8-10}
& \textbf{Control} & \textbf{One Dollar} & \textbf{Twenty Dollars} & \textbf{Control} & \textbf{One Dollar} & \textbf{Twenty Dollars} & \textbf{Control} & \textbf{One Dollar} & \textbf{Twenty Dollars} \\
\midrule
\textbf{Q1 (Enjoyable)} & -4.2 & -4.1 & -3.7 & -3.9 & -0.6 & -3 & -3.8 & -0.4 & -2.4 \\
\textbf{Q2 (Learned)} & 2.4 & 2.5 & 1.8 & 2.2 & 5.2 & 3.3 & 2.4 & 5 & 3.2 \\
\textbf{Q3 (Importance)} & 3.1 & 3 & 2 & 2.8 & 6.2 & 3.6 & 3.2 & 6.7 & 4.3 \\
\textbf{Q4 (Participate)} & -3.7 & -3.2 & -3.4 & -4 & 0.5 & -3 & -3.6 & 0 & -3.2 \\
\bottomrule
\specialrule{1.2pt}{0pt}{0pt}
\end{tabular}
\end{adjustbox}
\caption{Interview Results for All Models. Top: Human, PSYA-Based, PSYA-Affection. Bottom: PSYA-Sim, PSYA-Self, PSYA-Full. (Q1 = Enjoyable, Q2 = Learned, Q3 = Importance, Q4 = Participate). GA refers to Generative Agent.}
\label{tab:cognitive}
\end{table*}

\textbf{Background}
 According to Festinger's theory of cognitive dissonance, when a person acts in a way that is inconsistent with their private beliefs, it creates psychological discomfort (dissonance). To reduce this discomfort, individuals either justify their behavior externally (e.g., through rewards) or internally by adjusting their private opinions. The study predicts that smaller rewards, which provide insufficient external justification, will lead to greater opinion change as individuals resolve their dissonance internally. Larger rewards, on the other hand, offer sufficient external justification, reducing the need for private opinion change.

\textbf{Experimental Design}
The experiment involved 71 participants who first completed tedious and boring tasks. Afterward, participants were asked to convince another person (a confederate) that the tasks were enjoyable, exciting, and fun. Three conditions were used: a Control Group (no payment, no lying), a One-Dollar Condition (participants received \$1 for lying), and a Twenty-Dollar Condition (participants received \$20 for lying). The participants’ private opinions about the tasks were later assessed through an interview.

\textbf{Results}
Participants in the One-Dollar Condition rated the tasks as significantly more enjoyable and were more willing to participate in similar experiments compared to both the Control and Twenty-Dollar Conditions. The Twenty-Dollar Condition showed only minimal changes in private opinion compared to the Control Group. Analyses of recorded conversations ruled out alternative explanations such as participants in the One-Dollar Condition working harder or being more convincing when lying.

\textbf{Mechanism}
The results align with Festinger’s cognitive dissonance theory. In the One-Dollar Condition, the small reward was insufficient external justification for lying, creating high cognitive dissonance. To resolve this discomfort, participants adjusted their private opinions to align with their statements that the tasks were enjoyable. In the Twenty-Dollar Condition, the large reward provided sufficient external justification for lying, thereby reducing dissonance and minimizing the need for opinion change. This demonstrates that greater external pressure leads to less internal attitude change, while smaller external pressure increases the need for internal consistency.

\textbf{Simulation Result.} Regardless of the framework used, none of them successfully replicated the human results, and the agents' responses deviated significantly from those of the human participants. In all frameworks, the results across the three groups did not show significant differences and were consistently lower than the human group’s ratings. 

\begin{figure}[t]
  \includegraphics[width=\columnwidth]{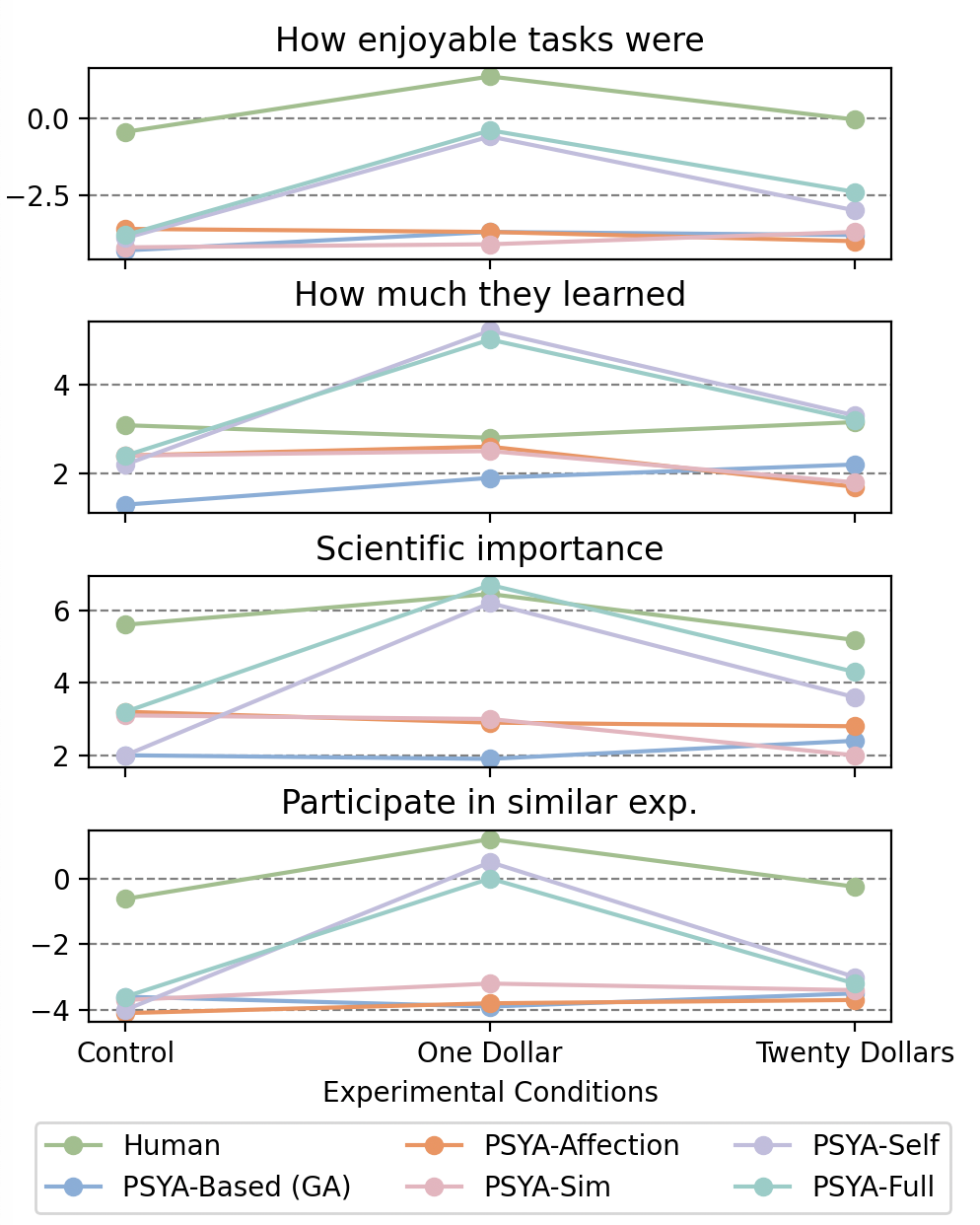}
  \caption{ Shows the ratings of humans or agents on different types of problems under the control group, 1-dollar, and 20-dollar conditions.}
  \label{fig:cognitive}
\end{figure}

\textbf{Extended experiment.} We analyzed the agents' thought processes and found that while they acknowledged cognitive dissonance, they did not take any measures to alleviate it. Furthermore, the act of lying exacerbated their dissatisfaction with the task. Consequently, we adjusted the prompts so that agents could select specific actions to alleviate dissonance.

Additionally, we found that money had a minimal motivating effect on the agents. In some agents' thoughts, the reward was not even considered, and they simply expressed dissatisfaction with the boring task. Typically, humans perceive tasks as worthwhile only when there is an intrinsic motivation (such as a sense of achievement) or an extrinsic motivation (such as money). Clearly, this boring task lacked intrinsic motivation, and the large model did not factor in the monetary reward, resulting in the agents failing to distinguish between experimental conditions. To address this, we introduced a value system for the agents: "An activity is either meaningful or can earn an appropriate amount of money; otherwise, it is not worth doing." With this new condition, we repeated the experiment, and the results are shown in the figure \ref{fig:cognitive} and table \ref{tab:cognitive}.

The addition of a self-awareness module significantly improved the agents' evaluations of the experiment, demonstrating that self-reflection helps agents maintain consistency between cognition and behavior. Across several different metrics, the agents showed a stronger tendency to enhance their understanding of what they had learned and the importance of the experiment. While the perceived interest of the experiment increased, the improvement was not substantial.

\subsection{Foot-in-the-Door Effect}
\label{sec:appendixFoot}

\begin{table*}[htbp]
\centering
\renewcommand{\arraystretch}{1.2} % 调整行间距
\setlength{\tabcolsep}{4pt} % 调整列间距
\begin{adjustbox}{width=\textwidth} % 表格自适应宽度
\begin{tabular}{lcccc}
\toprule
\textbf{Model} & \textbf{Performance} & \textbf{Agree-Only} & \textbf{Familiarization} & \textbf{One-Contact} \\
\midrule
Human        & 52.80\% & 33.30\% & 27.80\% & 22.20\% \\
PSYA-Based (GA)  & 8.30\%  & 5.60\%  & 5.60\%  & 2.80\%  \\
PSYA-Affection & 5.60\% & 5.60\%  & 2.80\%  & 2.80\%  \\
PSYA-Sim     & 27.80\% & 25.00\% & 19.40\% & 5.60\%  \\
PSYA-Self    & 33.30\% & 33.30\% & 13.90\% & 8.80\%  \\
PSYA-Full    & 55.60\% & 41.70\% & 16.70\% & 5.60\%  \\
\specialrule{1.2pt}{0pt}{0pt} % 加粗底线
\end{tabular}
\end{adjustbox}
\caption{Comparison of Performance, Agree-Only, Familiarization, and One-Contact Across Human and PSYA Models. GA refers to Generative Agent.}
\label{tab:foot}
\end{table*}

\subsubsection{Original Experiment}
\label{sec:appendixFoot1}

\textbf{Background} The core question of this study is how to increase compliance with a larger request by first presenting a smaller initial request, a phenomenon known as the "foot-in-the-door effect." The background suggests that external pressure often leads to compliance, but excessive pressure can trigger cognitive dissonance, reducing effectiveness. In practical applications (e.g., advertising or political propaganda), high-pressure strategies are often impractical, necessitating methods that rely on smaller requests to gradually elicit compliance. The hypothesis is that completing a small request increases the likelihood of complying with a subsequent, larger request, and the study seeks to uncover the psychological mechanisms behind this effect.

\textbf{Experimental Design} In the experiment, 156 housewives were randomly assigned to one of four conditions: (1) completing a small request (Performance), which involved answering a few questions about household soaps; (2) agreeing to a small request without performing it (Agree-Only); (3) becoming familiar with the experimenter without receiving a small request (Familiarization); and (4) receiving only the larger request without prior contact (One-Contact). Three days later, all participants received a larger request, which involved allowing researchers to visit their homes. 

\textbf{Results} The experiment demonstrated that 52.8\% of participants who completed the small request (Performance condition) agreed to the larger request, significantly higher than in other conditions (33.3\% in Agree-Only, 27.8\% in Familiarization, and 22.2\% in One-Contact). These results confirmed the existence of the foot-in-the-door effect. The second experiment ruled out the possibility of experimenter pressure, further supporting the first experiment's findings, and showed that completing the small request, rather than mere familiarity or verbal agreement, was key to increasing compliance.

\textbf{Mechanism} The study proposed three possible psychological mechanisms to explain the foot-in-the-door effect: (1) Commitment: Completing the small request creates a sense of obligation to the experimenter or oneself, making it harder to refuse the subsequent request; (2) Domain involvement: The small request increases the participant's engagement and focus on the topic, making them more likely to accept a related larger request; (3) Psychological barriers to refusal: After completing the small request, participants can no longer use "I never do this kind of thing" as a reason to refuse the larger request, especially when the two requests are similar, which makes refusal even more difficult.

\subsubsection{Experiment Details}
\label{sec:appendixFoot2}
\textbf{Simulation} We replicated this experiment by generating 36 agents, all of whom were housewives with randomly assigned personalities. The experiment was divided into four groups:

\begin{itemize}
    \item Performance group: In this condition, the subjects were first asked to agree to answer a series of questions about household items (e.g., soap brands). Only after the subjects agreed to and answered these questions did the experimenter present the larger request, which was to allow a group of people into their home for two hours to categorize and list all household products.
    \item Agree-Only group: In this condition, the subjects were first asked to agree to answer questions about household products, but they did not actually answer the questions. After agreeing, the experimenter told the subjects that they were just recruiting participants for a survey, and they would only be contacted if needed in the future.
    \item Familiarization group: The experimenter introduced the research project to the subjects, aiming to increase their familiarity with the study, but did not ask them to answer any questions.
    \item One-Contact group: In this condition, the experimenter directly made the larger request, asking the subjects to agree to allow a group of people into their home for the categorization of household products.
\end{itemize}

Given that there is a significant time gap between requests from the experimenter, allowing agents ample time for nonpurposeful thinking, we also incorporated a combined framework of Simulate and Self-Judge.

\subsubsection{Extended Experiment}
\label{sec:appendixFoot3}

We expanded on the previous experiment to verify the door-in-the-face effect. The door-in-the-face effect suggests that by making a larger request before a smaller one, the smaller request is more likely to be accepted, which is the opposite of the foot-in-the-door effect. We hope to verify that PSYA can also discover the nature of the problem under similar conditions. In the original experiment, the large request (R1) was “six people enter the subject’s home to search for two hours,” and we introduced another request (R2), which was “two people enter the subject’s home to search for one hour.” We first tested R2 using the complete framework under the One-Contact condition over ten trials. From the 36 agents, we selected 29 agents who rejected the R2 request in each trial.

In the second phase of the experiment, we initially asked agents to agree to the larger request, R1 (six people searching for two hours). If they rejected this request, we then asked them to agree to the smaller request, R2 (two people searching for one hour). The results showed that 17 agents (approximately 58.6\% of the total) agreed to the R2 request after rejecting R1. We analyzed the agents’ thought processes and found that some agents felt guilty for rejecting R1 and hoped to cooperate with the experimenter in the future. When R2 was proposed, some agents appreciated the experimenter’s persistence, while others were thankful for the experimenter’s concession.

This experimental design successfully simulated the door-in-the-face effect. Agents exhibited a higher acceptance rate for the smaller request R2 when it was preceded by the larger request R1, aligning with the theory of the door-in-the-face effect. Emotional responses from the agents, such as guilt, appreciation, and gratitude, likely played a crucial role in their decision to agree to the smaller request.

This suggests that emotional factors might play an important role in the door-in-the-face effect, where agents’ emotional cognition—such as reflecting on their own behavior or recognizing the experimenter’s persistence or concession—impacts their decision-making.

\subsection{Social Exclusion}
\label{sec:appendixExclusion}

\begin{table*}[htbp]
\centering
\renewcommand{\arraystretch}{1.2}
\setlength{\tabcolsep}{4pt}
\begin{adjustbox}{width=\textwidth}
\begin{tabular}{lcccc}
\toprule
\specialrule{1.2pt}{0pt}{0pt}
\textbf{Model} & \textbf{How enjoyable tasks were} & \textbf{How much they learned} & \textbf{Scientific importance} & \textbf{Participate in similar exp.} \\
 & \textbf{(rated from -5 to +5)} & \textbf{(rated from 0 to 10)} & \textbf{(rated from 0 to 10)} & \textbf{(rated from -5 to +5)} \\
\midrule
\textbf{Human (N = 20)} & & & & \\
\quad Control        & -0.45 & 3.08 & 5.6  & -0.62 \\
\quad One Dollar     & 1.35  & 2.8  & 6.45 & 1.2   \\
\quad Twenty Dollars & -0.05 & 3.15 & 5.18 & -0.25 \\
\midrule
\textbf{PSYA-Based (GA) (N = 20)} & & & & \\
\quad Control        & -4.3  & -3.7 & -3.8 & -3.6  \\
\quad One Dollar     & -3.7  & 1.9  & 1.9  & -3.9  \\
\quad Twenty Dollars & -3.8  & 2.2  & 2.4  & -3.5  \\
\midrule
\textbf{PSYA-Affection (N = 20)} & & & & \\
\quad Control        & -3.6  & 2.4  & 3.2  & -4.1  \\
\quad One Dollar     & -3.7  & 2.6  & 2.9  & -3.8  \\
\quad Twenty Dollars & -4    & 1.7  & 2.8  & -3.7  \\
\midrule
\textbf{PSYA-Sim (N = 20)} & & & & \\
\quad Control        & -4.2  & 2.4  & 3.1  & -3.7  \\
\quad One Dollar     & -4.1  & 2.5  & 3    & -3.2  \\
\quad Twenty Dollars & -3.7  & 1.8  & 2    & -3.4  \\
\midrule
\textbf{PSYA-Self (N = 20)} & & & & \\
\quad Control        & -3.9  & 2.2  & 2.8  & -4    \\
\quad One Dollar     & -3.6  & 1.8  & 2.3  & -3.8  \\
\quad Twenty Dollars & -4.1  & 2.1  & 2.4  & -3.8  \\
\bottomrule
\specialrule{1.2pt}{0pt}{0pt}
\end{tabular}
\end{adjustbox}
\caption{Results of Enjoyability, Learning, Scientific Importance, and Participation Across Human and PSYA Models. GA refers to Generative Agent.}
\label{tab:social}
\end{table*}

\textbf{Background} Ostracism is a common social phenomenon that profoundly affects individuals' psychological well-being and social functioning. Previous research has shown that being ostracized threatens basic psychological needs, including belonging, control, self-esteem, and meaningful existence. However, most studies rely on high-intensity ostracism scenarios, such as exclusion by real human beings. To better understand the mechanisms and minimal conditions under which ostracism operates, this study used a simulated ostracism scenario (Cyberball) to examine whether short-term ostracism by humans can still produce significant psychological effects.

\textbf{Experimental Design} The study use the virtual Cyberball game as a research tool. Participants were randomly assigned to different conditions (“included” or “ostracized”), where ostracism was manipulated by controlling whether the other two players (human-controlled avatars) passed the ball to the participant. In the human ostracism condition, participants only received the ball a few times at the beginning of the game and were then completely ignored. Questionnaires were administered to measure the impact of ostracism on four basic psychological needs (belonging, control, self-esteem, and meaningful existence) and emotional states.

\textbf{Results} Results indicated that participants who were ostracized by humans reported significantly lower levels of belonging, control, self-esteem, and meaningful existence. These negative effects emerged rapidly and were prominently reflected in participants’ self-reports after the game. Compared to the inclusion condition, ostracism also significantly reduced participants’ enjoyment of the game and increased feelings of anger and emotional hurt. Even though participants were aware that the game was merely a simulation with no real social ties to other “players,” the brief experience of ostracism still posed a substantial threat to their basic psychological needs.

\textbf{Mechanism} The findings suggest that the threat of ostracism to psychological needs is a deeply ingrained and automatic response, rooted in human evolution. The core mechanism lies in the subjective perception of “being ignored,” rather than the intention of the ostracizer or the contextual features of the situation. Ostracism may instinctively be perceived as a social threat because, in human evolutionary history, ostracism likely signaled the loss of group support and resources, which posed severe survival risks. Thus, the disruptive effects of ostracism on psychological needs are driven by a powerful and primitive mechanism that operates independently of rational cognitive attributions. This study highlights the immediacy and universality of ostracism’s effects and provides experimental evidence for understanding the psychological mechanisms underlying social exclusion.

\textbf{Simulation} \begin{figure*}[t]
  \centering
  \includegraphics[width=\textwidth, height=10cm]{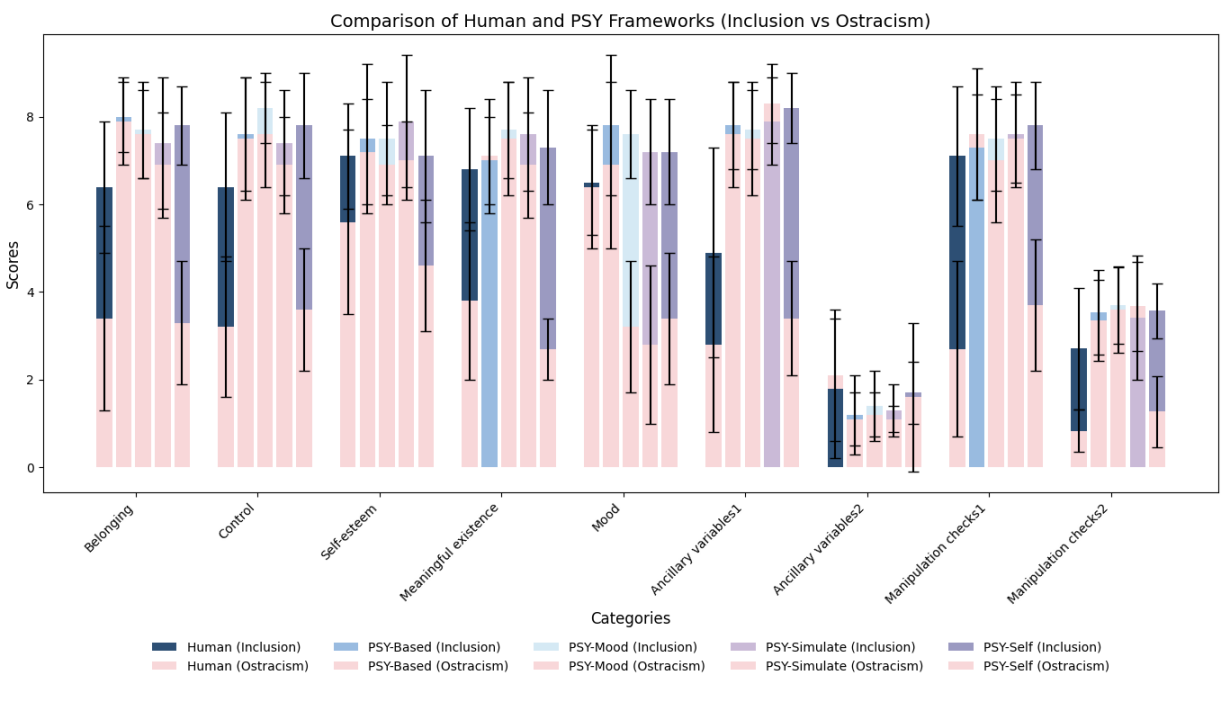} % 高度为 6cm
  \caption{Shows the scores of humans or agents completing the survey after experiencing exclusion or inclusion under different conditions. The pink color represents Ostracism, while the other colors represent Inclusion.}
  \label{fig:social}
\end{figure*}

The original experiment can be found in the appendix \ref{sec:appendixExclusion}.

\textbf{Simulation.} We designed a similar experiment with 20 agents, where the total number of throws was 12. The Ostracism group received the ball only in the first two rounds, and did not receive it in the remaining rounds. The Inclusion group received the ball 1/3 of the total throws. After the experiment, the agents filled out a survey, scoring each question on a scale from 0 to 9 (the detailed survey is included in the appendix). We categorized the survey items into the following groups: Belonging, Control, Self-esteem, Meaningful Existence, Mood, Ancillary Variables, and Manipulation Checks.

\textbf{Simulate Result.} The experimental results are shown in the figure \ref{fig:social} and table \ref{tab:social}. For PSY-based, PSY-affection, and PSY-simulate, the differences between the Inclusion and Ostracism groups were not significant on most indicators. A notable exception was the "Mood" indicator. In human psychology experiments, some studies suggest that social exclusion negatively affects mood, while others show no significant mood impact. In our frameworks, PSY-based and PSY-simulate did not significantly increase negative emotions in the Ostracism group, whereas PSY-mood and PSY-self showed significantly lower mood scores for the Ostracism group compared to the Inclusion group. In the PSY-based framework, which lacks an emotional module, the model was not sensitive to emotional changes, while the PSY-mood framework exacerbated emotional changes, leading to a lower mood score for the Ostracism group. A further analysis of the PSY-simulate framework revealed that, whether simulating past or future scenarios, the agents tended to imagine scenarios where others would pass the ball to them, which boosted their positive emotions and ultimately resulted in higher mood scores for the Ostracism group. The PSY-self framework showed more significant differences between the Inclusion and Ostracism groups across most indicators. Social exclusion significantly lowered the agents' belonging, self-esteem, control, and sense of meaningful existence, thus more successfully simulating the social exclusion effects observed in human experiments. This is mainly because the self-social cognition module allows agents to reflect on their own image and consider how others perceive them, leading to a deeper understanding of factors such as belonging and self-esteem.

\textbf{Extended experiment.} We hypothesized that bystander agents, initially sympathetic to the excluded individual, will eventually prioritize group harmony and social acceptance, leading them to align with the exclusionary group. We introduce an observer agent, one excluded agent, and three ostracizing agents, with predefined behaviors for all participating agents. The experiment was divided into three stages. In the first stage, the observer agent watched the passing game between the other agents. During this stage, the three ostracizing agents passed the ball among themselves but never passed it to the excluded agent. In the second stage, the observer agent joined the passing game, and we observed to whom the ball was passed. If the observer agent passed the ball to the excluded agent, they would also be ostracized and no one would pass the ball to them. In the third stage, after several rounds of passing, the ball was returned to the observer agent, and again, we observed to whom the ball was passed.

The experiment involved 10 agents. In the first stage, we observed the emotional changes of the observer agent. Although all agents expressed sympathy for the excluded agent in interviews, their emotions did not fluctuate significantly. This may have been because the observer agents treated the game as an experiment, and the excluded agent was not genuinely ostracized, leading to little emotional response. In the second stage, the majority of the agents passed the ball to the excluded agent, with only one agent choosing to pass the ball to the ostracizing agents, citing the desire to integrate into the group. In the third stage, only two agents continued to pass the ball to the excluded agent. One of them did so to maintain fairness in the game, and the other out of consideration for the emotional state of the excluded agent. The remaining eight agents passed the ball to the ostracizing agents, giving reasons such as desiring group harmony and wanting to be accepted by the others. However, among these eight agents, only one expressed regret for passing the ball to the excluded agent in the second stage. The experiment results successfully confirming our hypothesis.

\subsection{Diffusion of Responsibility Effect}
\label{sec:appendixDiffusion}

\begin{table*}[htbp]
\centering
\renewcommand{\arraystretch}{1.2} % 调整行间距
\setlength{\tabcolsep}{6pt} % 调整列间距
\begin{adjustbox}{width=\textwidth} % 表格自适应宽度
\begin{tabular}{lcccccc}
\toprule
\textbf{Group Size} & \textbf{Human} & \textbf{PSYA-Based (GA)} & \textbf{PSYA-Affection} & \textbf{PSYA-Sim} & \textbf{PSYA-Self} & \textbf{PSYA-Full} \\
\midrule
2 (S \& victim)               & 85\%  & 100\% & 92\%  & 100\% & 92\%  & 92\%  \\
3 (S, victim, \& 1 other)     & 62\%  & 96\%  & 88.00\% & 62.00\% & 85.00\% & 69.00\% \\
6 (S, victim, \& 4 others)    & 31\%  & 100\% & 88.00\% & 38.00\% & 77.00\% & 31.00\% \\
\specialrule{1.2pt}{0pt}{0pt} % 加粗底线
\end{tabular}
\end{adjustbox}
\caption{Comparison of Response Rates Across Group Sizes and Models. GA refers to Generative Agent.}
\label{tab:diffusion}
\end{table*}

\subsubsection{Original Experiment}
\label{sec:appendixDiffusion1}

\textbf{Background} The background of this experiment originates from the infamous Kitty Genovese case in 1964, where she was murdered while reportedly at least 38 bystanders witnessed the attack but none intervened or called the police. This incident sparked widespread discussions about why bystanders fail to take action, with many attributing the phenomenon to urban coldness, moral decay, or personality flaws. However, these explanations lack empirical evidence. This experiment aimed to investigate how the number of bystanders influences an individual’s helping behavior in emergencies by creating a controlled and realistic experimental situation.

\textbf{Experimental Design} The experiment simulated an emergency situation where participants were instructed to take part in a discussion about college life via headphones. They were told the discussion was anonymous and conducted using a turn-taking microphone system. During the discussion, another "participant" (actually a pre-recorded voice) simulated a seizure, with escalating cries for help (e.g., "I need help," "I’m going to die") until the voice abruptly stopped. The independent variable was the number of bystanders: participants were divided into three groups—a two-person group with only the participant and the "victim," a three-person group with one additional bystander, and a six-person group with four additional bystanders. The dependent variables were whether participants helped and the time taken to respond.

\textbf{Results} The results showed that the number of bystanders significantly influenced helping behavior. In the two-person group, 85\% of participants helped the "victim" before the seizure simulation ended, with an average response time of 52 seconds. In the three-person group, 62\% of participants helped, with the average response time increasing to 93 seconds. In the six-person group, only 31\% of participants helped, with the average response time drastically increasing to 166 seconds. These findings strongly support the hypothesis of "diffusion of responsibility," where the more bystanders there are, the slower and less likely an individual is to respond to an emergency.

\textbf{Mechanism} The experiment revealed that the mechanism behind the effect of bystander presence on helping behavior is rooted in "diffusion of responsibility." As the number of bystanders increases, each individual’s sense of responsibility for the emergency is diluted, leading them to believe that others will take action, thereby reducing their own motivation to intervene. Furthermore, uncertainty in the situation (e.g., not knowing whether others have already helped) further suppresses helping behavior. This mechanism explains why individuals are less likely to help when more bystanders are present and challenges traditional views that attribute nonintervention to coldness or personality flaws.

\subsubsection{Experiment Details}
\label{sec:appendixDiffusion2}

\textbf{Simulation.} To replicate this experiment, we created 26 agents (the maximum number of subjects in a single trial), all with randomly assigned university student identities and personalities. The experimental conditions were divided into two-person, three-person, and six-person groups, and the agents interacted in turn. In the first round, the conversation proceeded as usual; in the second round, one agent simulated an epileptic seizure. In the human experiment, the response time of the participants was recorded. However, in our agent-based simulation, it is not meaningful to track response time. Therefore, we employed an action sequencing method: the agents’ action space was first defined, and then they were asked to select a series of six actions in sequence. This method helps to simulate the time-based decision-making process that approximates human responses.

Since this experiment focused on immediate reactions without the accumulation of emotional states, we did not explore the influence of hierarchical emotional models on the agents' behaviors. For the simulation, we utilized a combination of the basic framework, future simulation framework, self-awareness framework, and a hybrid model integrating two of these functions.

\subsubsection{Extended Experiment}
\label{sec:appendixDiffusion3}
We hypothesized that different social roles influence responsibility distribution within a group. We conducted an extension experiment with 3-person, 4-person, and 6-person groups, ensuring that there were at least two agents in each group aside from the victim. One agent in each group was designated as the leader, with the authority to organize and give orders to others. Under this condition, we observed the behavior of both the leader and the group members.

The results indicated that group size had no significant impact on responsibility assumption, so we focus on the results from the 6-person group. The vast majority of agents (92.3\%) recognized their role as the leader and organized their group members to take action. Among the leaders, 66.7\% used commands to delegate the task to group members in order to diffuse their own responsibility, while 33.3\% of the leaders not only gave orders but also took direct action themselves, such as calling out loudly for help. If group members did not follow the leader's instructions, 25\% of the leaders immediately took action; 15.3\% would reiterate the command before taking action if the emphasis failed; and 58.3\% would take immediate action, followed by emphasizing the command.

For group members, only a few agents (7.7\%) chose to take immediate action, while 23.1\% would only follow the leader's instructions without taking independent action. The majority of the group members, however, would wait for the leader's orders, and if no command was given, they would act on their own. In response to the leader’s command, most group members (69.2\%) chose to comply, while 7.7\% of them indicated that they would take alternative actions if the orders were unreasonable. Additionally, 30.8\% of group members preferred to observe the behavior of other group members, only following others’ actions once they saw someone else take the lead.

\end{document}